\documentclass[useAMS,usenatbib]{mn2e}
\usepackage{graphicx}
\usepackage{amssymb}
\usepackage{natbib}
\newcommand{\msun}{{M$_\odot$}}

\title{Slow Cooling in Low Metallicity Clouds: An Origin of Globular Cluster Bimodality?}

\author[R. Fernandez et al.]{Ricardo Fernandez$^{1}$ and Greg L. Bryan$^{1,2}$\\
$^{1}$Department of Astronomy, Columbia University, 550 West 120th Street, New York, NY 10027 \\
$^{2}$Center for Computational Astrophysics, Flatiron Institute, 162 Fifth Avenue, New York, NY 10010}

\begin{document}

\date{}

%\pagerange{\pageref{firstpage}--\pageref{lastpage}} \pubyear{2013}

\maketitle

%\label{firstpage}

\begin{abstract}

We explore the relative role of small-scale fragmentation and global collapse in low-metallicity clouds, pointing out that in such clouds the cooling time may be longer than the dynamical time, allowing the cloud to collapse globally before it can fragment.  This, we suggest, may help to explain the formation of the low-metallicity globular cluster population, since such dense stellar systems need a large amount of gas to be collected in a small region (without significant feedback during the collapse).  To explore this further, we carry out numerical simulations of low-metallicity Bonner-Ebert stable gas clouds, demonstrating that there exists a critical metallicity (between 0.001 and 0.01 $Z_\odot$) below which the cloud collapses globally without fragmentation.  We also run simulations including a background radiative heating source, showing that this can also produce clouds that do not fragment, and that the critical metallicity -- which can exceed the no-radiation case -- increases with the heating rate.

\end{abstract}

\begin{keywords}
globular clusters - methods:numerical
\end{keywords}

%% ----------------------------------------------------------------
%
\section{Introduction}

Globular clusters (GCs), with typical masses of $10^5$-$10^6$ \msun\ are particularly interesting relics of star formation for a number of reasons, including the following: (1) the are very concentrated, with half-mass radii of a few parsecs, indicating that star formation occurred in a particularly dense environment; (2) the stars in a given GC generally have a very narrow spread in age and metallicity, implying a single stellar population (although recent results have revealed a more nuanced situation here, as we will discuss briefly later), and (3) the metallicity distribution of GCs in external galaxies is generally bimodal (or at least very different from the metallicity distribution of stars in the galaxy as a whole), with a large number of low-metallicity GCs.  Reviews of their properties include \citet{Brodie2006}, \citet{Renzini2008, Renzini2013}, \citet{Kruijssen2014Review}, and see also \cite{zwart2010}.

This bimodal, or possibly skewed metallicity distribution \citep[e.g.,][]{Gebhardt1999, Strader2003, Peng2006} is sometimes interpreted as indicating that there are two formation modes: one that produced low-metallicity, old systems and a second for the generally younger, higher-metallicity component.  For instance, \citet{Ashman1992} suggested that metal-rich GCs are formed in gas rich mergers and metal-poor GCs are donated by progenitor spirals.  However, their work did not incorporate a cosmological model, and their predictions of the number and colour distribution of GCs in massive Es galaxies were not consistent, as pointed out by \citet{Forbes1997}.  

\citet{Beasley2002} augmented this picture by incorporating a semi-analytical model of combined galaxy and GC formation in a cosmological context. In that work, each mode of GC formation was assigned a fixed efficiency relative to the field stars. However, to match observed values, the formation of metal-poor GCs had to be artificially truncated after $z = 5$.  Later work explored other aspects of the cosmological context: for example \citet{Prieto2008} modeled the evolution of an initially power-law high-redshift metal-poor GC population under the time-varying gravitational potential expected in cosmological galaxy evolution.  \citet{Boley2009} argued that disruption of these blue GCs could significantly contribute to the galactic halo population.  \citet{Gray2011} suggested that GCs could form in low-mass halos enriched by galactic outflows.  The age difference between metal rich and metal poor populations \citep{Hansen2013}, as well as the spatial distributions of globular clusters can also inform their origin \citep{Hargis2014}.  Other models explored reionization for setting the bimodality \citep[e.g.,][]{Santos2003, Harris1994}  \citep[although see][]{Forbes2015}.

Recently, investigations have explored more empirical, hierarchical galaxy formation models in a cosmological setting to explain the bimodal metallicity distribution.   \citet{Muratov2010} followed the formation of GCs using the assembly history from cosmological simulations combined with observed scaling relations. In their model, bimodality naturally arises from the rate of galaxy mergers. Early mergers preferentially produce metal-poor GCs and a few late massive mergers can produce a significant number of metal-rich GCs.  This was extended to more massive halos in \citet{Li2014} and incorporated in full cosmological simulations in \citet{Li2016} and \citet{Renaud2017}.     \citet{Tonini2013} also found that the metal bimodality could be reproduced based on an observationally fixed mass-metallicity relation and mass-GC formation efficiency.  

A difficult challenge for connecting formation models to the observed present-day population is GC destruction and mass-loss by both internal and external influences \citep{Fujii2007, Kruijssen2012}.  Indeed, \citet{Lamers2017} argue that the metallicity dependance of the cluster specific frequency is largely due to varying destruction efficiencies.  Another way to connect the formation of globular clusters to the rest of the stars in the galaxy was explored by \citet{Kruijssen2015}, who developed models for the formation and survival of high-mass clusters as a natural part of star formation, with an efficiency that depends on interstellar medium properties (and hence cosmology).

While they included the cosmological picture, essentially none of the works discussed above attempted to model the detailed structure of star formation within collapsing proto-globular clusters.  In particular, it is not clear how to get a large amount of gas ($10^6$ \msun) into a very small region without star formation occurring during the collapse, which would result in a wide spatial distribution of stars and perhaps even prevent the collapse due to feedback.  In this work, we explore the collapse and fragmentation of gas clouds in low-metallicity environments.

As an aside, we note that, quite recently, observations have demonstrated that globular clusters are not a single stellar population, but may be composed of multiple generations showing enhanced He and specific abundance changes, particularly those associated with proton-capture processes \citep[e.g.,][]{Norris1981, Kraft1994, Gratton2001, Carretta2009}.  In addition, photometric data shows a splitting of the main sequence in many GCs \citet[e.g.,][]{Piotto2009, Anderson2009, Milone2010}.  This has been challenging to explain because, with a few possible exceptions, the Fe abundance distribution is generally very narrow (consistent with observational errors), indicating that supernova self-enrichment plays no role.  A wide range of models have been proposed to explain these abundance irregularities, beginning with the possibility that AGB stars in the 4-8 \msun\ mass range can produce the necessary elements through hot bottom burning \citep[e.g.,][]{DErcole2010, Ventura2013}.  Other ideas include the existence of Fast Rotating Massive Stars \citep[FRMS][]{Krause2013}, supermassive stars \citep{Denissenkov2014,  Denissenkov2015}, and massive interacting binaries \citep[e.g.,][]{deMink2009, Bastian2013}.  All of these solutions are problematic for a number of reasons \citep[e.g.,][]{Renzini2015, Bastian2015}, including the mass budget required to generate the observed number of second generation stars.  However, in this work, we will not explicitly explore this second generation, instead focusing on the general problem of understanding fragmentation and collapse in low-metallicity gas. Indeed, although we have discussed this in the context of globular cluster formation, we are really trying to understand how low-metallicity gas cools and collapses.

%% ----------------------------------------------------------------
%
\subsection{Basic Idea}
\label{sec:basic}

In this paper, we explore a simple idea: can the cooling properties of low-metallicity gas clouds themselves influence how star formation proceeds?  Higher-metallicity (by which we mean approximately solar metallicity, or even lower -- we will address this point more precisely below) gas cools rapidly, typically on a timescale shorter than the dynamical time, meaning that present day large gas clouds, with masses in the GC range, are typically ``cold", with thermal temperatures well below their virial temperatures and so rapid fragmentation is inevitable \citep{Hoyle1953}.  This generally means that solar metallicity giant molecular clouds will rapidly produce stars before they are completely collapsed and feedback from those stars will result in a low star formation efficiency \citep[e.g.,][]{McKee1989}.  However, for a low enough metallicity, the gas may cool slowly so that the cloud will collapse coherently, not fragmenting until the central gas density is very high.  These high densities promote rapid star formation resulting in high efficiency.  In this way, paradoxically, low efficiency cooling may result in high efficiency star formation. 

To further investigate this simple idea we have created a very simple model of a cooling parcel of gas which compares the relative importance of gravity and cooling. In this picture, the parameter space consists of density, temperature, and metallicity. Once the parameters are chosen the evolutionary timescales are computed: the quantity of interest is the ratio of the absolute value of the cooling time to the dynamical time $t_{\rm cool}|/t_{\rm dyn}$, where $t_{\rm dyn} = \sqrt{3 \pi / 32 G \rho}$. In our simple model, cooling is computed using the publicly available \textsc{grackle} chemistry library.  To do this, we use the tabulated, equilibrium mode of \textsc{grackle} \citep{Smith2016}.  This cooling rate is based in turn on an equilibrium calculation from Cloudy \citep{Cloudy2017} (this does not include molecular cooling, but here we are primarily interested in general trends -- see section~\ref{sec:numerical_method} for more discussion).

\begin{figure}
\begin{center}
\mbox{\includegraphics[width=8.7cm]{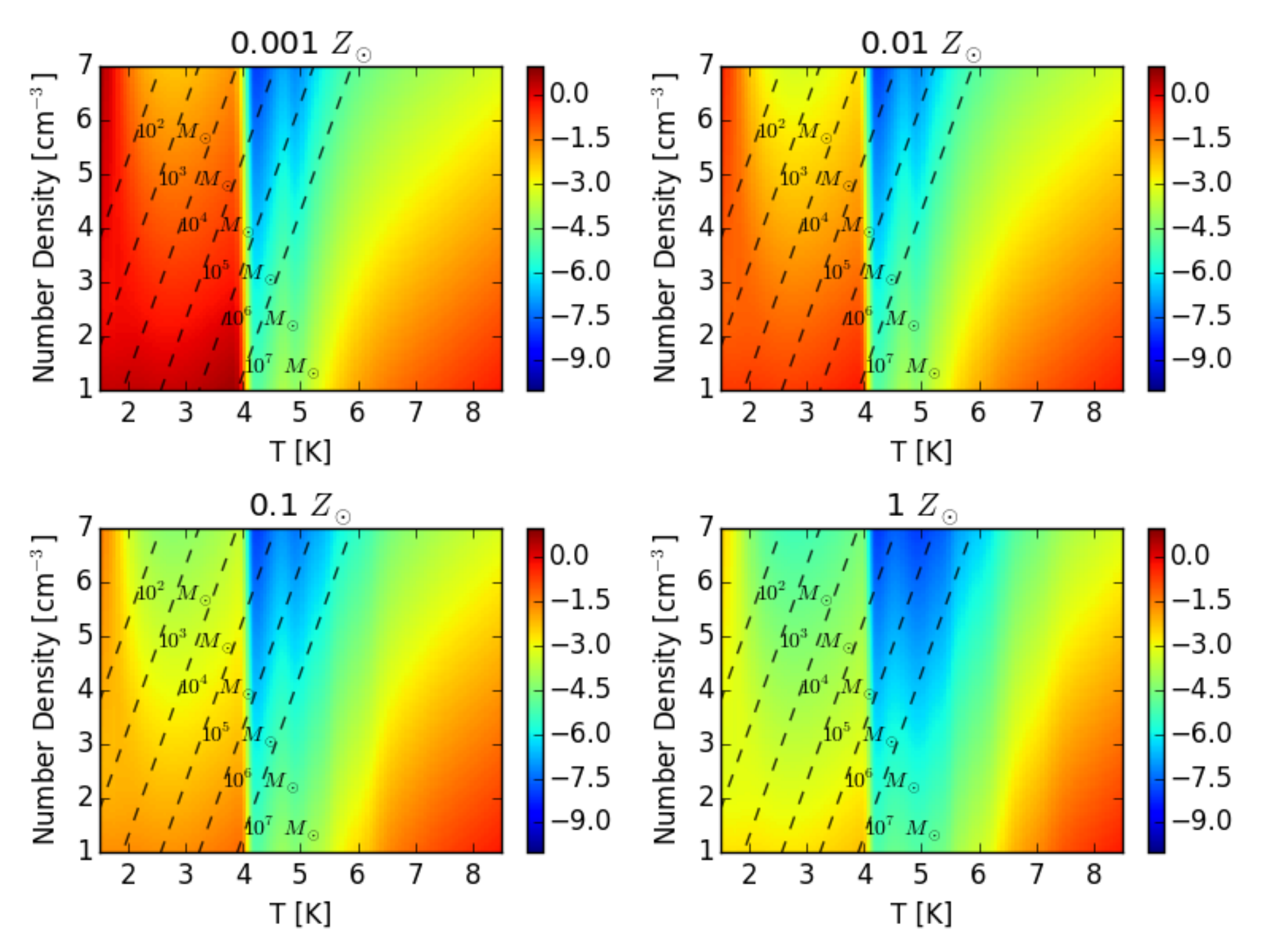}}
\end{center}
\caption{\label{fig:cooling_to_freefall} The ratio of cooling time to dynamical time for gas with a range of density, temperature and metallicity values, as labelled.}
\end{figure}

Figure~\ref{fig:cooling_to_freefall} is a panel showing the ratio of log$(|t_{cool}|/t_{dyn})$ for metallicity values $Z/Z_{\odot}=10^{-3},10^{-2},10^{-1},1$.  To begin, we focus on the lowest metallicity  (upper-left) panel of Figure~\ref{fig:cooling_to_freefall} -- we see that there are two regions where cooling and dynamical time are comparable (shaded red). The first is in the temperature range $10^1-10^4$K (the sharp cutoff at $10^4$K is due to efficient cooling from HI line emission), and the second is in the bottom right corner. It is the former region in which we are most interested because this area is where we expect to find gas clouds with values conducive to globular cluster formation. Over each heat map, we plot lines of density and temperature corresponding to constant Bonner-Ebert mass (see Section~\ref{sec:numerical} for details on how this is computed). It is clear from the plot that, for gas clouds with masses typical of globular clusters, there are values in the region where cooling is relatively inefficient, allowing the gas to collapse coherently before it can cool and fragment. Further, as the metallicity increases, this region becomes less pronounced and the gas becomes more efficient in cooling, which will allow the gas to cool and fragment before global collapse sets in.

\begin{figure}
\begin{center}
\mbox{\includegraphics[width=8.7cm]{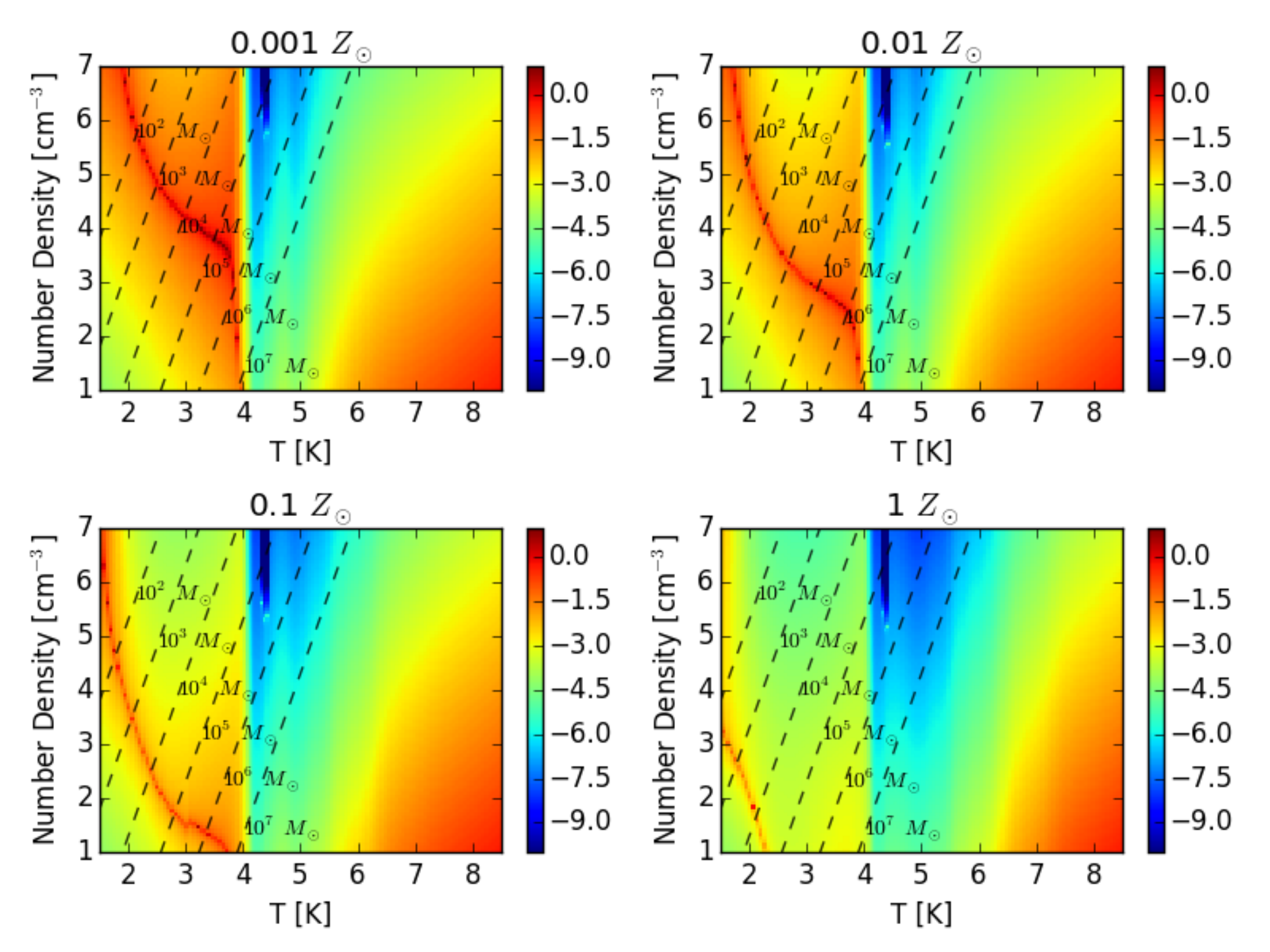}}
\end{center}
\caption{\label{fig:cooling_to_freefall_background} The ratio of cooling time to dynamical time for gas with a range of density, temperature and metallicity values, as labelled.  This plot is similar to Figure~\ref{fig:cooling_to_freefall}, except that we include a radiative heating source as specified in \citet{Haardt2012}, at $z=$ 0.}
\end{figure}

This is all computed in the absence of any radiative background.  In Figure~\ref{fig:cooling_to_freefall_background}, 
the one zone models are recalculated as previous except allowing for radiative heating.   Although the radiative background
is uncertain, we adopt the radiative background from \citet{Haardt2012} at $z=$ 0 
in order to show the kind of effect we expect.  This does not change the high gas cooling rate, but it does affect the lower-temperature gas cooling time.   In particular, we see that an equilibrium curve is present where
heating and cooling are balanced. We no longer have the extended region where cooling and dynamical time scales are 
comparable but now are concentrated along the equilibrium curve. Furthermore, it should be expected that gas will
naturally seek the equilibrium curve values.  Moreover, when the
the metallicity increases, the equilibrium curve shifts downward, decreasing the possibility of having globular cluster
like conditions.

This is all determined by computing cooling and dynamical times for gas with a characteristic density and temperature, demonstrating that the idea of inefficient cooling may be appropriate for low-metallicity gas (or gas with a somewhat higher metallicity but stronger radiative background).   We now turn to space- and time-dependent numerical simulations to explore this idea further.  Ideally, we would carry out cosmological simulations that included the full range of dynamical processes relevant for star formation at high-redshift with low (but non-zero) metallicity.  However, this is computationally intractable, and therefore we instead investigate a simple, idealized set up.  We expect that gas cloud collisions during mergers at high-redshift will result in the accumulation of gas in relatively dense clouds.  These clouds will rapidly cool to temperatures around $10^4$ K.  Therefore, we set up turbulently perturbed Bonner-Ebert spheres with masses typical of globular clusters, and densities/metallicities motivated by Figure~\ref{fig:cooling_to_freefall}.   In future work, we will explore more complicated dynamics, such as colliding flows; however, here we explore perhaps the most simple possible test of this idea.

Finally, we connect this work to research on the formation of the first generation of stars out of a completely primordial gas which are thought to be quite massive, and their transition to regular star formation \citep[for reviews see][]{Bromm2011, Glover2013}.  This transition is typically assumed to occur at a `critical' metallicity, at which cooling due to metals is more effective than cooling due to molecular hydrogen \citep{Bromm2001}, although the precise value and meaning of the critical metallicity is not uniformly agreed upon.  \citet{Omukai2000} explored one-zone models following the chemical and thermal evolution of clouds with a variety of metals, first drawing attention to the way in which gas at low metallicity cools.  Later work \citep{Bromm2003, Omukai2005, Santori2006} explored this in more detail, generally finding that, at high density, this critical metallicity was around $10^{-4}$ to $10^{-3}$ Z$_{\odot}$ (but, depending on definition, was higher at lower densities).  The current paper is related to this work, in that we are very interested in fragmentation in collapsing clouds, but we explore fragmentation in somewhat different conditions.  In particular, we assume that the metallicity is always larger than this `critical` metallicity such that the stars that ultimately form are typical of the Population II initial mass function (IMF) and so include the low mass stars seen in globular clusters.  Instead, we examine when fragmentation can be delayed during the collapse process, to produce rapid star formation at high density.

%% ----------------------------------------------------------------
%
\section{Numerical Models}
\label{sec:numerical}
\subsection{Numerical Method}
\label{sec:numerical_method}

The simulations in this paper were performed with the publicly available Eulerian three-dimensional
hydrodynamical adaptive mesh refinement Enzo code \citep{Bryan2014}. The domain
box size of the simulation was 150 pc on a side with a top level root grid resolution of $128^3$, two
additional levels of initial refinement, and a maximum refinement level of 4, for a minimum cell size of 0.073 pc.  Cell refinement was dictated 
by the gas mass such that a cell was refined whenever its mass became larger than 0.1 M$_\odot$.
In addition, we refined based on the Jeans length such that it was always refined by at least 4 cells (up until
the maximum refinement is reached).

Our simulations include self gravity as well as radiative cooling using the
\textsc{grackle} library; details on cooling methods and assumptions are 
described in \cite{Smith2016}, but we summarize them briefly here.  The cooling (and
heating) rates are computed using a non-equilibrium model for H, H$^+$, He, He$^+$, He$^{++}$ and e$^-$,
while a look-up table in density and temperature is used for metal-line cooling (and heating), 
as described in \citet{Smith2016}, using Cloudy \citep{Cloudy2017}, based on the assumption of ionization equilibrium.  
For simplicity we use solar abundances, scaled to the (lower) adopted metallicity.
When a radiative background is included, we use \cite{Haardt2012} at $z=0$ (which is included in both
the non-equilibrium calculations for the primordial species and the equilibrium Cloudy 
calculations for the metal cooling).  For reference, this radiative background produces an equilibrium
temperature a factor of few below the the local interstellar radiation field \citep{Richings2014}.
The runs described below are all initialized with relatively
low electron fractions ($\sim 10^{-4}$) -- at such low values, fine-structure metal cooling is dominated
by collisions with hydrogen but higher cooling rates would arise if the ionization fractions were much higher.

We do not include molecular
cooling; given our assumed metallicity, this is reasonable for CO, but is potentially more problematic
for H$_2$, as emphasized recently by \citet{Glover2014}, who found that H$_2$ cooling could be an important
coolant at the densities of interest in these studies for metallicities below 0.1 Z$_\odot$.  In particular, they
demonstrated that for low and moderate UV backgrounds (photons in the Lyman-Werner bands being the
most important), H$_2$ cooling could reduce the gas temperature below $10^4$ K for densities in the 0.1 - 10$^4$ 
cm$^{-3}$ range.  However, for sufficiently high UV intensities,
molecular dissociation was effective; therefore, our neglect of H$_2$ is equivalent to saying that we are
considering only regions close to a bright UV source (as indeed, our simulations with radiative heating
implicitly also assume).  Future work extending this to a more complete treatment of molecular and
line cooling is planned.

%\begin{figure}
%\begin{center}
%\includegraphics[width=4cm]{Images/Initial_number_density}
%\includegraphics[width=4cm]{Images/Initial_temperature}
%\end{center}
%\caption{\label{fig:initial_setup} Slices showing the number density and temperature 
%for the initial conditions used in this paper. }
%\end{figure}

\subsection{Initial Conditions}

Our initial conditions consist of a cloud in pressure equilibrium with a
constant ambient density and temperature background. The internal structure of the
cloud is modeled by a Bonner-Ebert sphere \citep{Bonnor1956}: a self-gravitating
isothermal gas sphere in hydrostatic equilibrium embedded in a pressurized  
medium. To fully describe a Bonner-Ebert sphere, a mass $M_{BE}$, temperature
$T_{BE}$, and an external pressure $P_{ext}$ must be chosen. Following our
assumptions outlined in Section~\ref{sec:basic}, we chose
$M_{BE}=10^6$ \msun, $T_{BE}=6000$ K, and $P_{ext}/k_B=1.8\times10^5$ K cm$^{-3}$, 
where $k_B$ is the Boltzmann constant. This corresponds to a cloud on the point of
gravitational collapse with nearly comparable cooling and collapse times,
depending on the chosen metallicity and external heating (which we will vary). Table~\ref{table:parameters}
summarizes the parameters for each simulation, numbered by the order discussed in the paper.

\begin{table}
       \begin{center}
	\begin{tabular}{@{}ccccc}
	\hline
	%Run & $M$ (M$_{\odot}$) & $\overline{\rho}$ $(\mathrm{g/cm^{-3}})$ & $T$ (K) & $Z$ (Z$_{\odot}$) & $\Gamma$ ($\mathrm{erg/s/g}$)\\ 
	Run & $M$  & $T$ & $Z$  &  $\Gamma$\\ 
          & (M$_{\odot}$) & (K) & (Z$_{\odot}$) & (erg s$^{-1}$cm$^{-3}$)\\ 
	\hline
	    %1 & $10^6$ & 7.80$ \times 10^3$ & 6$ \times 10^3$ & $10^{-3}$ \\
        %2 & $10^6$ & 1.05$ \times 10^4$ & 6$ \times 10^3$ & $10^{-2}$ \\
        %3 & $10^6$ & 8.50$ \times 10^3$ & 6$ \times 10^3$ & $10^{-2}$ \\ 
        %4 & $10^6$ & 8.50$ \times 10^3$ & 6$ \times 10^3$ & $10^{-2}$ & 8$ \times 10^{-26}$\\ 
        %5 & $10^6$ & 8.50$ \times 10^3$ & 6$ \times 10^3$ & $10^{-2}$ & 8$ \times 10^{-26}$\\ 
	    1 & $10^6$ & 6$ \times 10^3$ & $10^{-3}$ &  \\
        2 & $10^6$ & 6$ \times 10^3$ & $10^{-2}$ &  \\
        3 & $10^6$ & 6$ \times 10^3$ & $10^{-2}$ &  HM2012 \\ 
        4 & $10^6$ & 6$ \times 10^3$ & $10^{-2}$ &  HM2012 + 8.5$ \times 10^{-26}$\\ 
        5 & 5$\times 10^6$ & 6$ \times 10^3$ & $10^{-2}$ & HM2012 + 8.5$ \times 10^{-25}$\\ 
       \hline
	\end{tabular}
	\caption{Parameters for each simulation numbered by order of discussion in the paper.  After the simulation number, columns are: cloud mass, initial temperature, metallicity, and heating rate ($\Gamma$).  Heating, if present, is computed using \textsc{grackle}, with the \citet{Haardt2012} rate, possibly plus a constant rate.}
	\label{table:parameters}
	\end{center}
\end{table}

Such a cloud may naturally arise during the early formation history ($z \sim 5$ to $10$)
of typical galactic halos \citep{Kim2017}.  The virial temperatures of such progenitor halos are
typically around $T_{\rm vir} \sim 10^{5}$ K, and the required densities of our cloud
correspond to roughly an overdensity of $10^4$ relative to the mean density at that epoch.
Therefore our required densities are modestly above typical halo values and require
only a small amount of cooling and compression.  We imagine that during the cosmological
evolution of such halos, gas is shock heated to the virial temperature and overdensities
of order $\sim 1000$ and then cools rapidly to $10^4$ K, with a corresponding increase in
the density.  At this point, it's evolution will stall (as made clear by the large ratios of $t_{\rm cool}/
t_{\rm dyn}$ seen in Figure~1).  The recombination time of such gas is short and so it is
natural for the gas to be largely neutral.  A more complete treatment of this part of the
evolution would be interesting and beyond the scope of this paper.
  Finally, we do not discuss the fate of any dark matter 
in this scenario, but note that our clouds are most likely to form during the merger
phase of such halos, when dynamics can lead to shell crossing and so the clouds may
naturally form away from the central dark matter peak.

In addition, we add turbulence to the cloud following a power spectrum of
$v_k^2 \propto k^{-4}$ for the velocity field.  We include only modes between
$k_{\rm min}=$ 9 and $k_{\rm max} =$ 19 (in units of the fundamental mode of
the cloud) such that the input modes are reasonably well resolved
and yet have wavelengths smaller than the cloud radius.  We set the turbulent velocities
such that the rms velocity of the gas is equal to the sound speed of the cloud $c_s = 7.97$ km/s.
Although somewhat arbitrary, these conditions, corresponding to a Mach number
somewhat below unity, are not atypical of expected conditions \citep{Li2015}.  Moreover, because of
the importance of pressure support, the turbulence here is primarily playing the role
of introducing some set of initial perturbations, quite different from cold clouds, 
in which the turbulent forcing is much more important is setting
the properties of the clumps that form.  Again, this is due to the inefficiency of cooling
in low metallicity clouds.
We use these initial conditions for essentially all of the runs analyzed in this paper.

%% ----------------------------------------------------------------
% 

\begin{figure}
\begin{center}
\includegraphics[width=9.5cm]{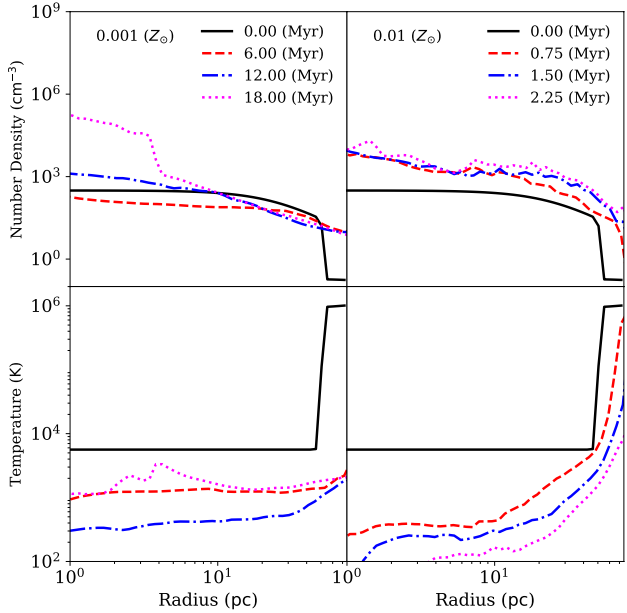}
\end{center}
\caption{\label{fig:profiles} Cell mass weighted profiles for 
density (top panel) and temperature (bottom panel) as a function of radius at various 
output times, as shown, for runs with cooling, turbulence, and metallicity of $Z=10^{-3}Z_\odot$
(left column) and $Z=10^{-2}Z_\odot$ (right column).}
\end{figure}

\section{Results}
\label{sec:results}

We now carry out a set of simulations exploring the evolution of this cloud under a variety of conditions, with a particular emphasis on the impact of metallicity on their evolution.  We begin, for simplicity, with models without any radiative background.

% ------

\subsection{No Heating Runs}

\subsubsection{$Z=10^{-3}Z_\odot$}

\begin{figure*}
\begin{center}
\hspace{-0.5cm}
\includegraphics[width=9.5cm]{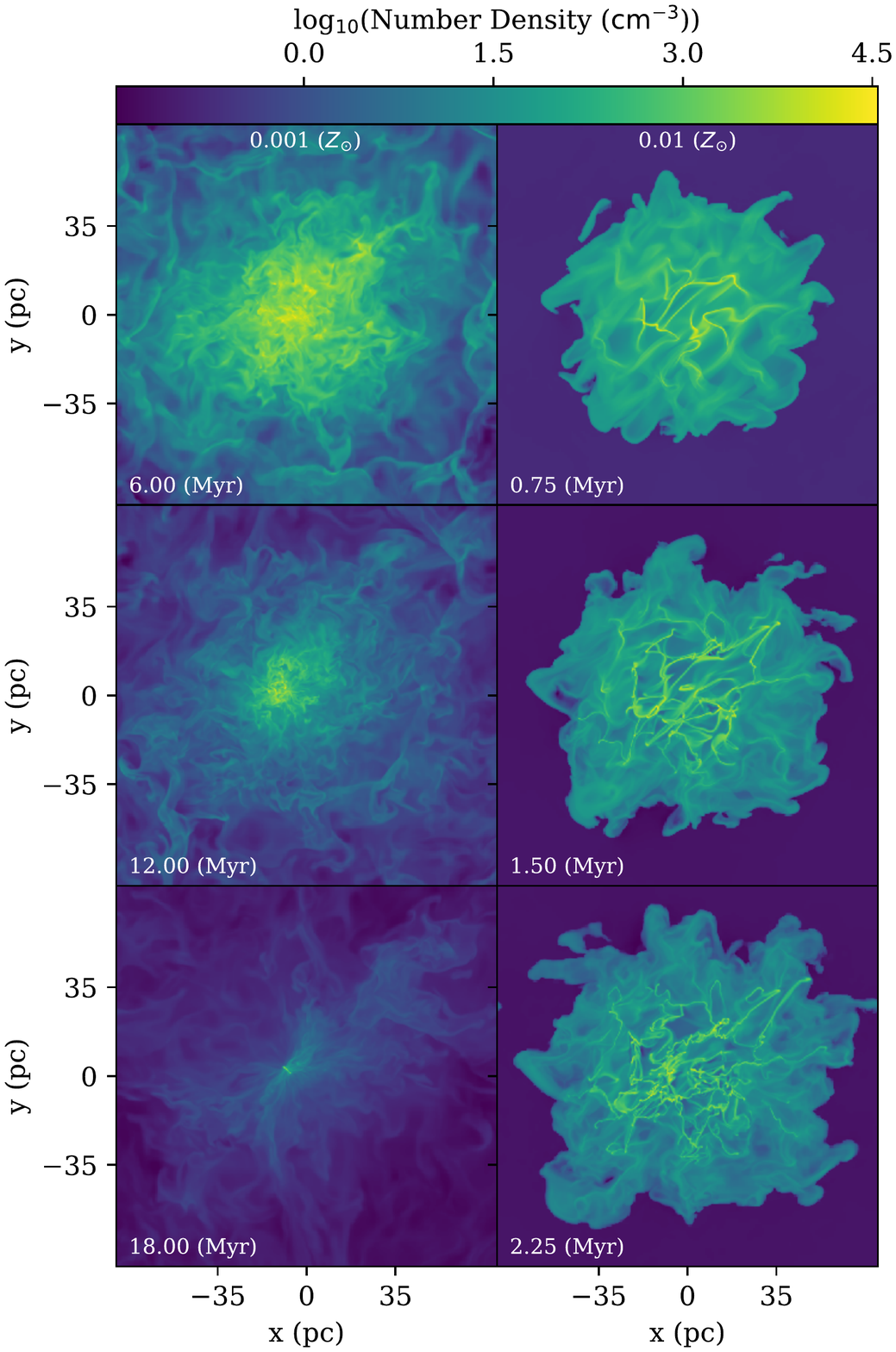} \hspace{-1.1cm}
\includegraphics[width=9.5cm]{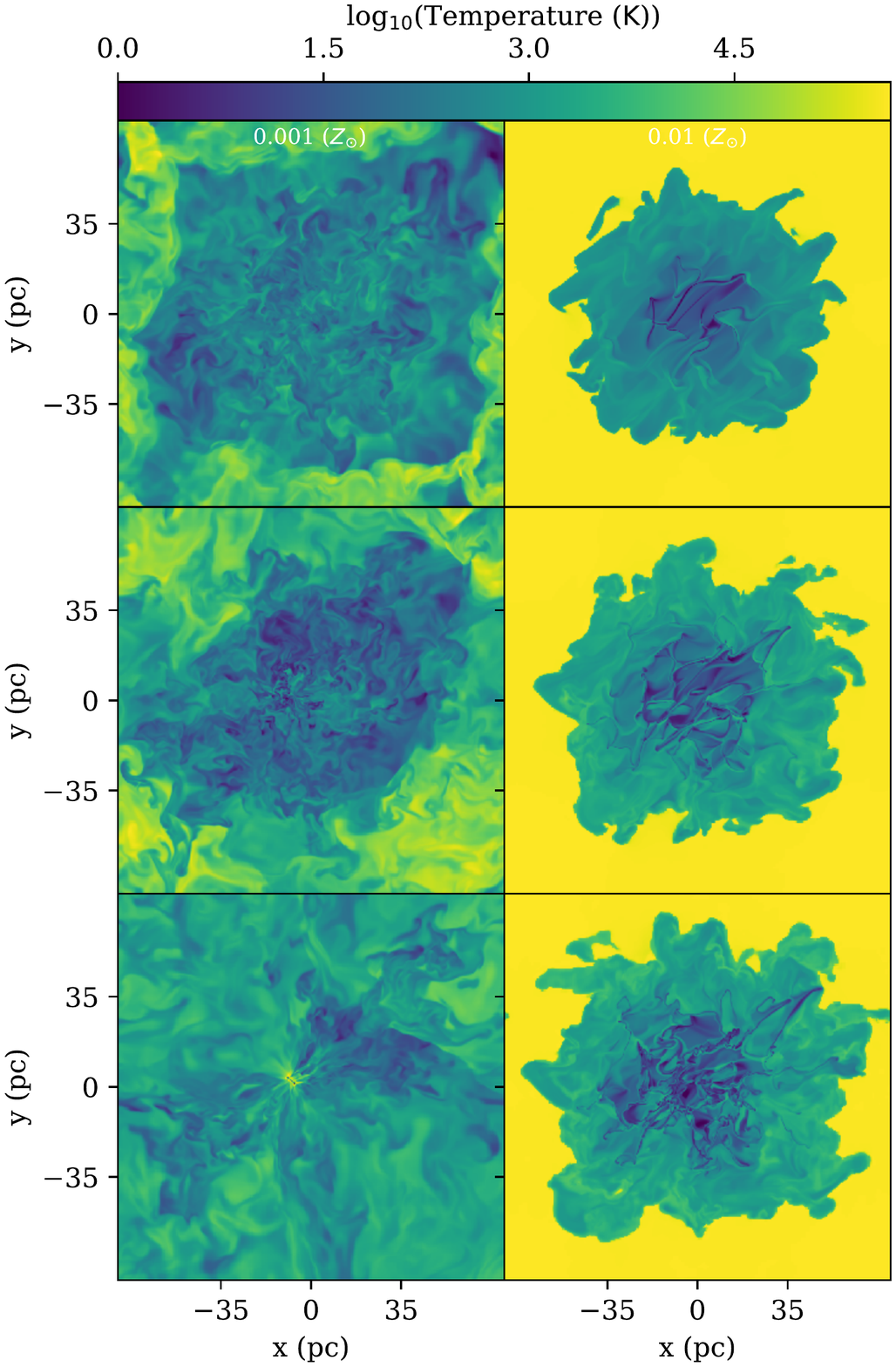} 
\end{center}
\caption{\label{fig:number_density_panel} Density (left) and temperature (right) slices of the evolution of the sphere without radiative heating.  Time evolution is from top to bottom and in each set of 6 panels, the left side is for the low metallicity ($Z=10^{-3}Z_\odot$) run and the right-side side is for the higher ($Z=10^{-2}Z_\odot$) run.  The times are the same times as shown in the profiles in Figure~\ref{fig:profiles}.  }
\end{figure*}

We start with a low metallicity gas -- adopting $Z=10^{-3}Z_\odot$ puts us well into the regime
where the gas cooling time is longer than the gravitational collapse time (see Section~\ref{sec:basic}).
In the left panels of Figure~\ref{fig:profiles}, we show density and temperature profiles at a range of times during
the collapse, stopping when high densities are reached and we can no longer accurately follow the evolution
(the Jeans length criteria cannot be met even at our highest allowed refinement level).  
The cloud, which is initially stable (i.e. in pressure equilibrium), starts to evolve due to both the added
turbulence and to gravity plus cooling. The free fall time of the cloud is $t_{ff}\approx 3$ Myr.
However, the slow cooling time delays the immediate large scale collapse.
This can be seen by the flat density profiles at times earlier than about 15 Myr. 
In fact, the cloud initially expands due to the added turbulence. The outer rim
of the cloud moves outward, briefly decreasing the density in the centre.
The expansion lasts for approximately 10 Myr, and during this time the temperature drops moderately
(by about a factor of 2), in part due to the expansion.  By 18 Myr, the gravitational
collapse sets in and a dense central core forms.   The bottom left-hand panel of Figure~\ref{fig:profiles}
shows the temperature profiles, with the temperature rising mildly during the recollapse, but
not heating above about 1000 K due to radiative cooling.

More detail of this collapse can be seen in Figure~\ref{fig:number_density_panel}, which shows slices of density (left set of panels) and temperature (right set of panels) for this low metallicity run on the left side of each set of panels.  We select the same times as in Figure~\ref{fig:profiles} (6, 12 and 18 Myr after the initial time).  Clearly the turbulence drives substantial fluctuations in the density (and temperature), but the cloud does not fragment, undergoing global collapse.   As expected from the one zone model, the cloud cannot efficiently cool before global gravitational collapse sets in.

% ---------

\subsubsection{$Z=10^{-2}Z_\odot$}

We repeat the previous run except we increase the metallicity to $Z=10^{-2}Z_\odot$.  The profiles are shown
in the right-hand side of Figure~\ref{fig:profiles} and the density/temperature slices on the right-hand side of the panels in Figure~\ref{fig:number_density_panel}.  In this case the evolution is very different.
The gas can now cool efficiently, as evident particularly in the temperature profiles.   The centre
of the gas cloud, up to a radius of about 10 pc, has cooled to $\approx 200$ K in less than a million years. The added effect of the turbulence
allows the cold gas to condense into dense pockets.  This is seen clearly in the density and temperature slices, which show the formation of many dense, self-gravitating filaments and clumps.  
Hence, we see again that our numerical runs agree with our simple one zone models. Moreover, we find that there is a critical metallicity
between $10^{-3}$ to $10^{-2}Z_\odot$ that separates the evolution of the gas cloud into either global gravitational collapse or local fragmentation. 

Note that the times shown in the profiles and slices differ between the two runs because we stop the calculation in both cases when dense gas clouds form and we are no longer able to follow the evolution even with our AMR run.  In each run, at this point, star formation would rapidly occur and so we stop the calculation.

\begin{figure}
\begin{center}
\includegraphics[width=9.0cm]{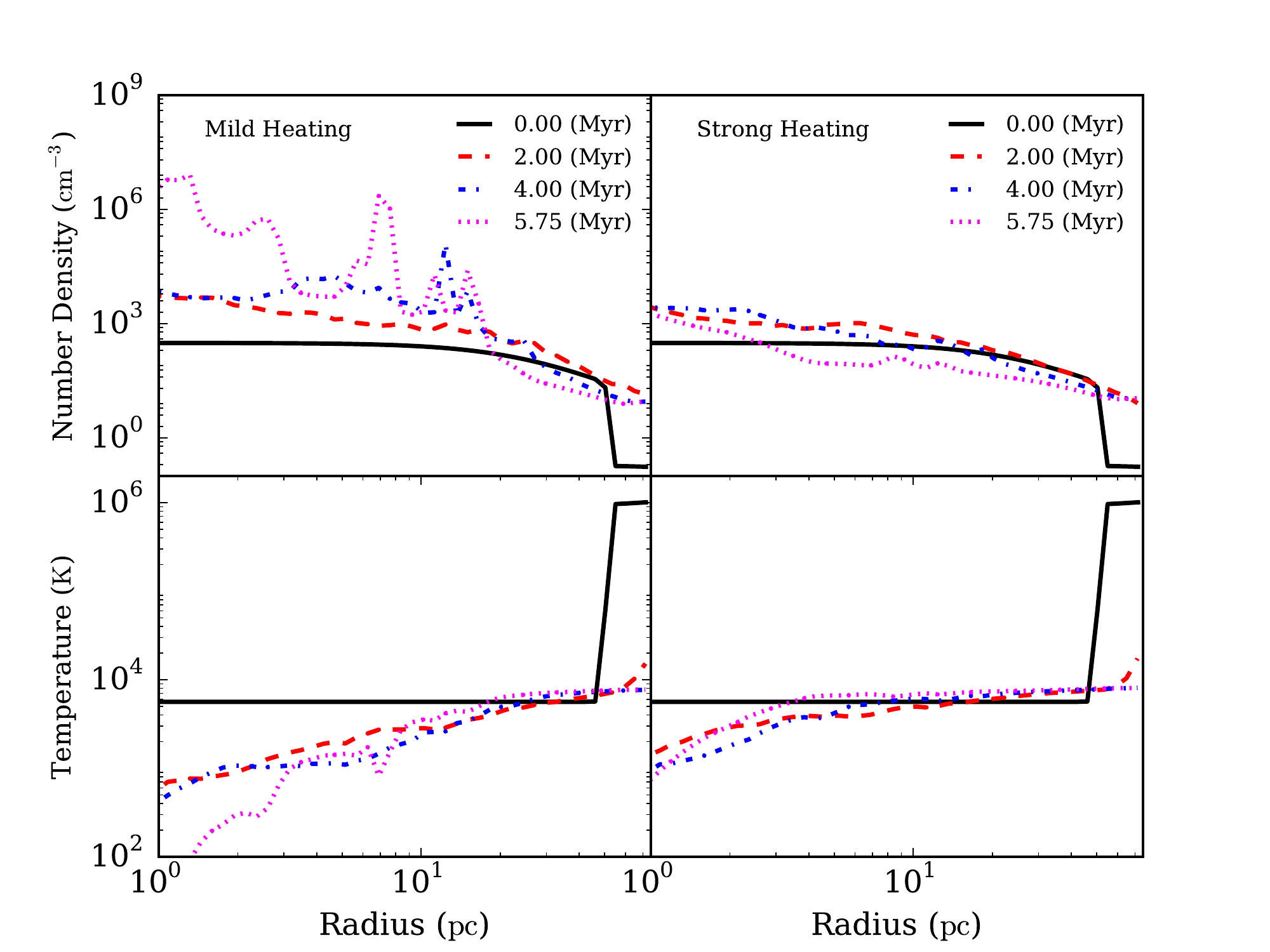}
\end{center}
\caption{\label{fig:profiles_heating} Cell mass weighted profiles for 
density (top panel) and temperature (bottom panel) as a function of radius at various 
output times, as shown, for runs with cooling, turbulence, and metallicity of $Z=10^{-2}Z_\odot$
and mild heating (left column) and stronger heating (right column).  See text for heating rates.}
\end{figure}

\begin{figure}
\begin{center}
\includegraphics[width=10cm]{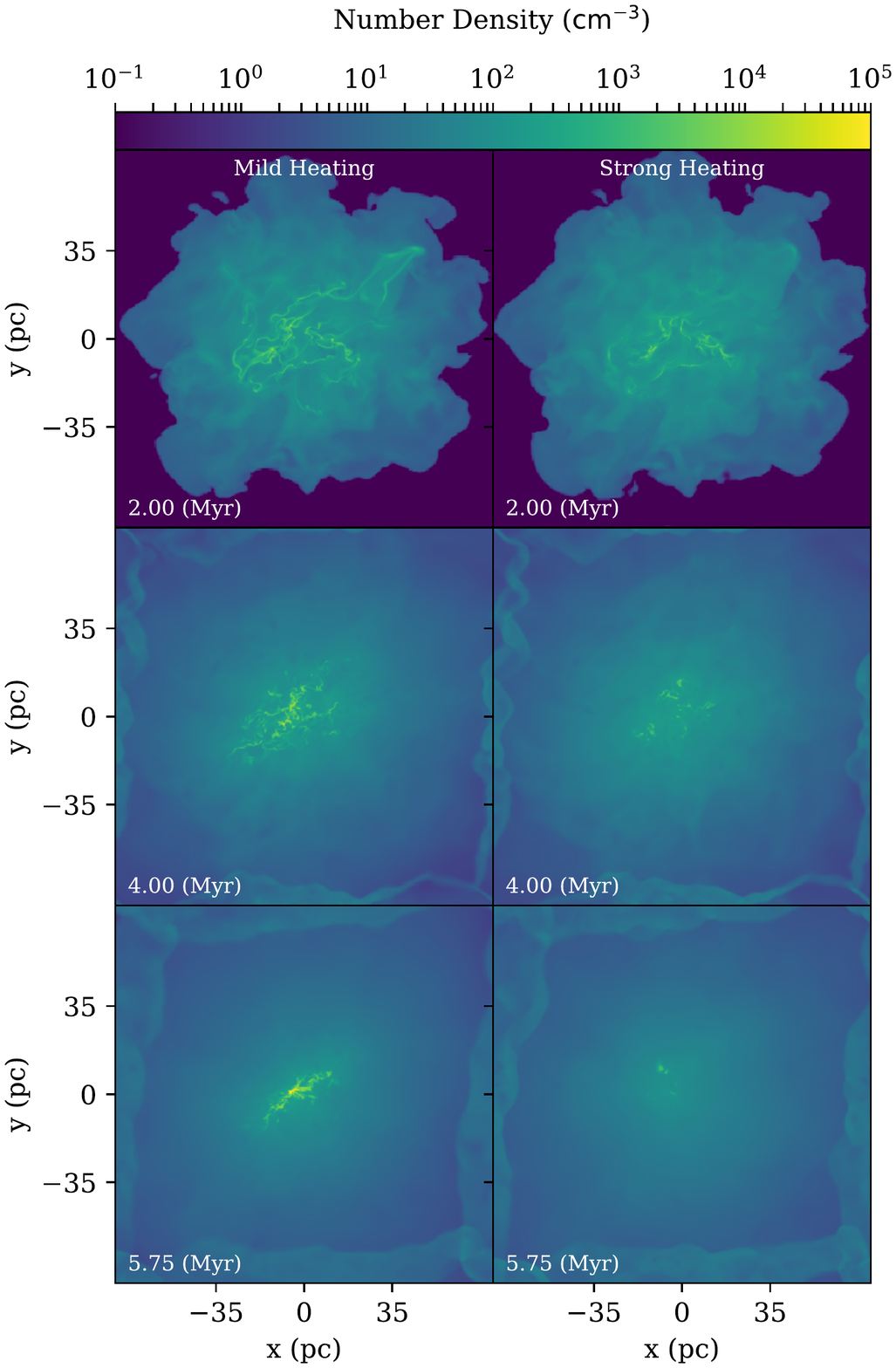}
\end{center}
\caption{\label{fig:density_heating}
Density slices at three different output times (as in Figure~\ref{fig:profiles_heating}) going
from top to bottom for the two different heating, rates: mild heating (left column) and
stronger heating (right column), all for $Z=10^{-2}Z_\odot$.
}
\end{figure}

% ------

\subsection{Runs with Photo-heating}

\subsubsection{$Z=10^{-2}Z_\odot$, varying heating}

We next turn to simulations which include the impact of radiative heating, either from a metagalactic background or from nearby star
formation (but assuming there is no direct physical impact on our cloud).  We have carried out two additions simulations to explore
this: the first has a radiative background typical of the \citet{Haardt2012} metagalactic background at $z=0$, as modeled with our
cooling package \textsc{grackle} (and ultimately Cloudy).  This is a relatively mild heating source (and also roughly equivalent to the UV background at $z \approx 6$), but is sufficient to heat the low-temperature gas at low metallicity (see Figure~\ref{fig:cooling_to_freefall_background} for the resulting cooling times).
The second simulation adds an additional specific heating rate of $8.25\times10^{-26}$ erg cm$^{-3}$ s$^{-1}$, as might arise from a photo-electric
heating term coming from a nearby unattenuated UV source (the requirements for which will be discussed in more detail later).

Figure~\ref{fig:profiles_heating} shows the radial density and temperature profiles for the two runs (left and right set of panels),
and Figure~\ref{fig:density_heating} shows the density slices for the same set of times and heating rates, also for the two runs
(left -- low heating, right -- higher heating). These plots demonstrate that a radiative heating source can substantially change
the nature of the collapse, for a fixed metallicity ($Z=10^{-2}$ $Z_\odot$ in this case).   Before discussing the results in more detail,
we note that the later two times (t = 4.0 and 5.75 Myr) show a shock around the edges of the box due to the (artificial) periodic
boundary conditions used.   These shocks arise from a mild outflow driven by the expansion of the cloud -- here caused mostly by the photo-heating --
and do not affect the evolution in the central part of the cloud.

\begin{figure*}
\begin{center}
\hspace{-0.4cm}
\includegraphics[width=9.4cm]{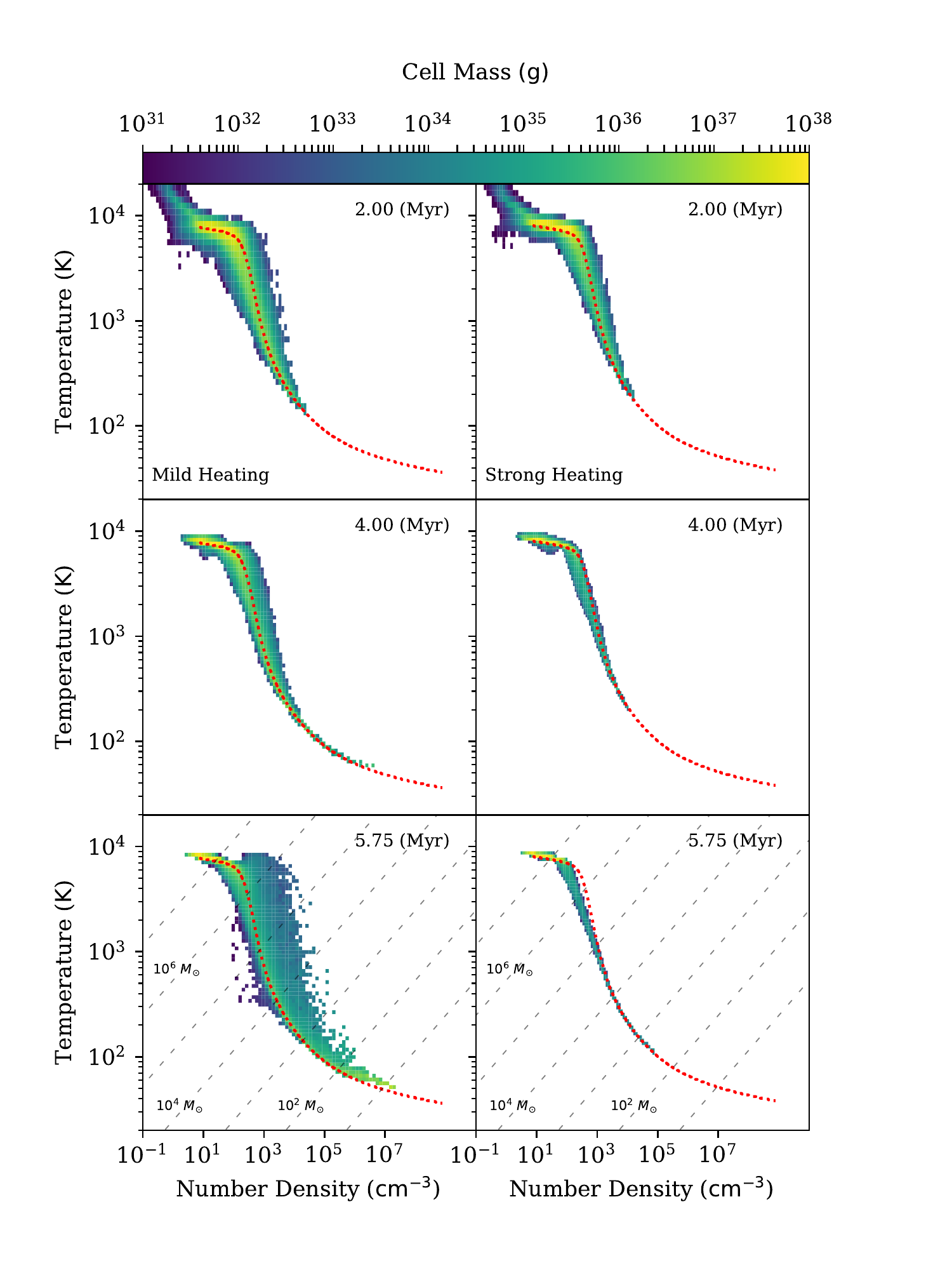} \hspace{-1cm}
\includegraphics[width=9.4cm]{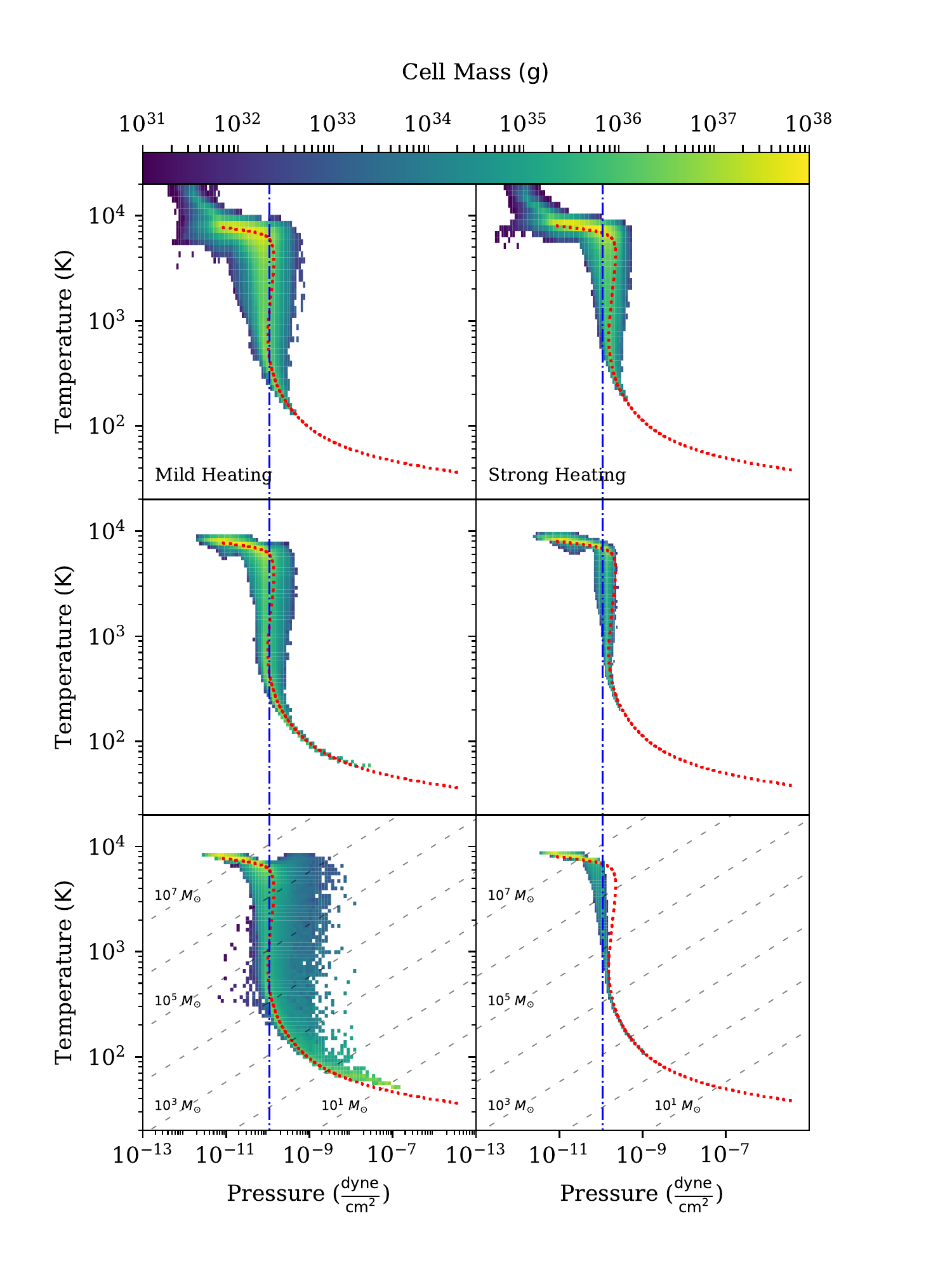} \hspace{-1.5cm}
\end{center}
\caption{\label{fig:phase_panels} Phase plots showing, in the left set of panels, the density-temperature distribution and, in the right set of panels, the pressure-temperature of the gas in our two simulations with varying radiative heating rates.   In each set of panels, time runs from top to bottom at the same time slices as in Figure~\ref{fig:density_heating}, and within each set of 6 panels the left side is for the low-radiative heating run, while the right side is the higher heating simulation.  The colour coding indicates the amount of mass in each phase.   In each plot, red dotted lines show the equilibrium density-temperature relation, while dot-dashed lines in the bottom row provides lines of constant Jeans mass, as labelled.  Finally, the vertical blue dot-dashed line is the Bonner-Ebert pressure defined in equation~\ref{eq:PcBE}.}
\end{figure*}

In the mild heating case, collapse proceeds in a roughly similar fashion to the no-heating run, except that the temperature does not fall as quickly, particularly in the centre. As the profiles and slices show, gas begins to fragment, but radiative heating prevents collapse to very high densities. Gas flows in to the centre and fragmentation eventually sets in there, with densities climbing to large values in the central 5-10 pc. The cloud ends up with a central multi-phase core, with forming clumps, and a smooth outer envelope. The core is not quite as compact as for the $Z=10^{-3}$ $Z_\odot$, no-radiation run, but compared to the $Z=10^{-2}$ $Z_\odot$, no-radiation case, fragmentation only occurs in the central 5-10 pc, rather than throughout the cloud.

The higher radiative heating case shows, despite only a relatively small increase in the heating rate, a substantially different evolution. Again, gas flows in and fragmentation begins in the central region, but this time the gas is unable to complete fragmentation, and the cloud ends up nearly entirely smooth with (almost) no star forming regions.

The reason behind this remarkable change in the evolution is easier to understand through phase diagrams.  In Figure~\ref{fig:phase_panels}, we show the density-temperature and pressure-temperature distributions of our two runs at the same times as in the previous plots.  The gas starts at $t=0$ (not shown) mostly in a small region in the left-hand side of each diagram (with $T \sim 6000$ K and $P \sim 10^{-10}$ dyne cm$^{-2}$).  In each case, the gas rapidly moves into thermal equilibrium with radiation and so follows the equilibrium density-temperature relation shown as a dotted line in the plots. The relation is particularly tight for the higher radiative heating run at late times, but is generally well-followed in both runs at all times. However, the larger difference between the two runs is the location along this curve to which the gas evolves;  at early times, in both cases, the gas is mostly in the warmer, lower-density phase. In the lower-heating run, some gas manages to cool and move into the lower right-region (where the Jeans mass decreases and star formation can commence), while in the higher heating run, essentially no gas moves in that direction.  

The reason for this difference is perhaps easiest to see in the pressure-temperature distribution plot.  Recall that the cloud begins and mostly evolves in pressure equilibrium, with a relatively small range of pressures.  Therefore, the cloud has two constraints in this diagram: the equilibrium curve as shown, and a nearly constant pressure constraint, coming from pressure equilibrium in the cloud.  We show this second constraint schematically as a line at the Bonner-Ebert pressure (define in equation~\ref{eq:PcBE}, below, although in detail there is some pressure variation from small to large radii).  This produces two stable regions: one at low temperatures ($T \sim 1000$ K and below) and one at higher temperatures, close to $10^4$~K.\footnote{Note that, although there is a temperature range (from about 1000-5000 K) over which the gas has multiple solutions at fixed pressure, the key point is that the pressure-temperature relation is very steep, note that it is formally multivalued.}

In the low heating run, there is some gas in the lower temperature phase, and this gas would like to cool and move up the equilibrium curve to the left, but in order to do so, it must dynamically increase its pressure, which it can only do through gravitational collapse (to do so, the amount of gas in this phase must exceed the Jeans mass, and it takes some time to collect together in the centre of the cloud).  In the higher heating simulation, there is essentially no gas in the lower phase, even though the extra heating is so small that it is difficult to see the small upward shift in the equilibrium curves in Figure~\ref{fig:phase_panels}.  This small extra heating boosts the equilibrium pressure of the low-temperature stable phase above the cloud pressure and so it is inaccessible to the gas.  The net result is a completely stable cloud, with no collapse or fragmentation.

Therefore we see that heating modifies the collapse by forcing the gas into one of two phases: cold or warm, and only the low temperature phase can gravitationally collapse.  This is reminiscent of the situation in the interstellar medium at the present day; however, here our cooling rates are much lower over all and so the densities are sufficiently high that the dynamical (gravitational) timescale of the overall ($\sim 10^6$ \msun) cloud becomes important.

One natural question is the ultimate fate of the gas which is able to collapse.  We suspect that if the metallicity of the gas were very low, below $10^{-4}$ or $10^{-5} Z_\odot$ (i.e truly primordial gas), this collapse would continue without substantial fragmentation, producing massive Population III stars \citep[e.g.,][]{ABN2002}.  For the higher metallicities considered here, fragmentation into lower mass stars is the natural outcome -- overplotted as dot-dashed lines in Figure~\ref{fig:phase_panels} are lines of constant Jeans mass.  It is clear that any gas in the cold and dense phase (in the lower right of the lower panels) is rapidly moving to lower Jeans mass, and so will ultimately fragment into very small clumps (hence the supposition of a standard Population II IMF for these stars seems reasonable).  The evolution of gas at these clumps has been investigated with, for example, the inclusion of dust cooling \citep{Clark2008, Dopcke2011, Dopcke2013}, demonstrating that low mass stars with metallicities as low as $10^{-5}$ Z$_{\sun}$.

In the discussion, we will develop a simple analytic model based on the insights described above and discuss how this model can be applied to physical conditions in order to create a more comprehensive picture for cloud collapse vs. fragmentation as a function of metallicity and radiative heating rate.  However, first, we will explore one final model which features a more extreme case.

\begin{figure*}
\begin{center}
\includegraphics[width=8.5cm]{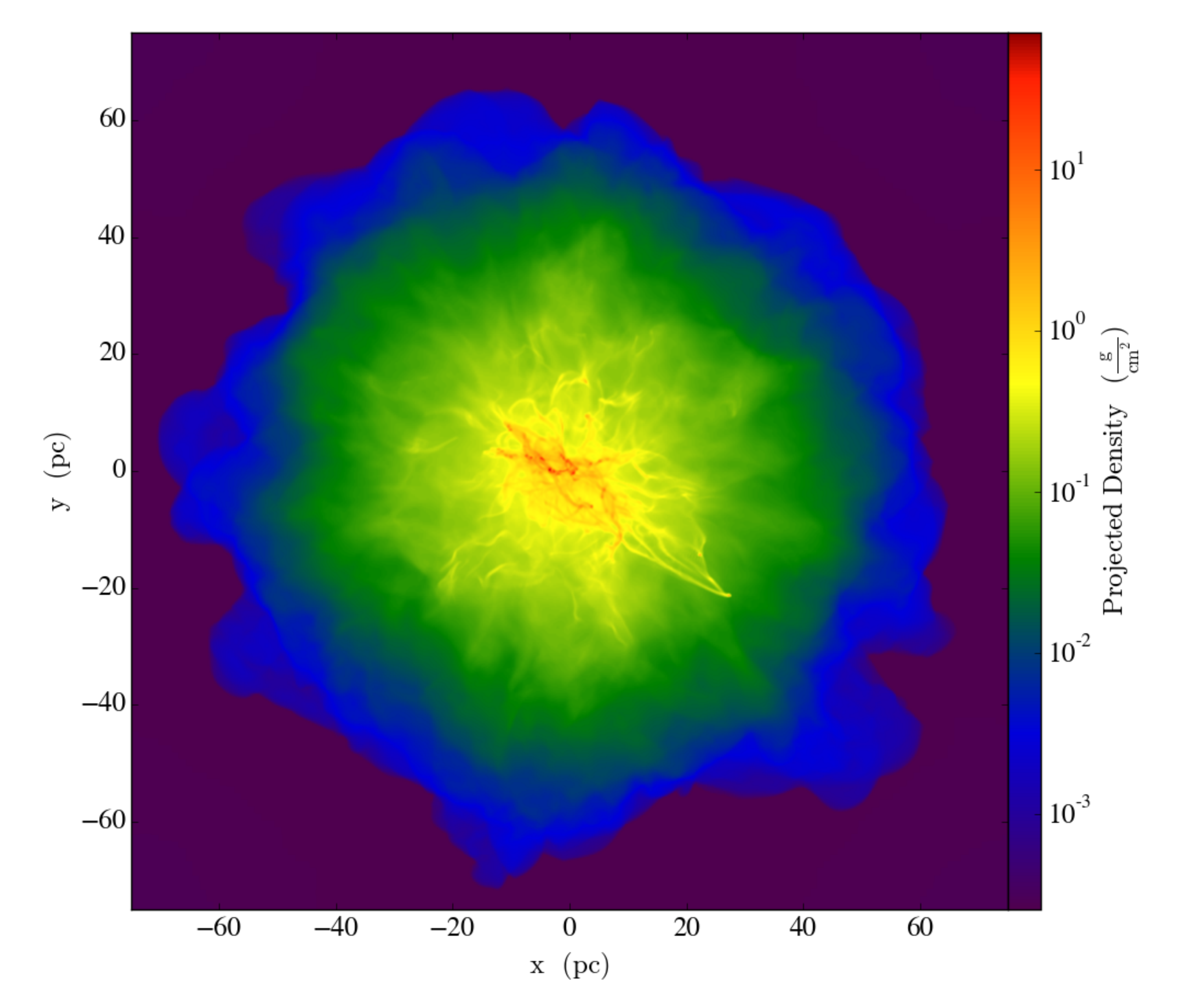}
\includegraphics[width=8.5cm]{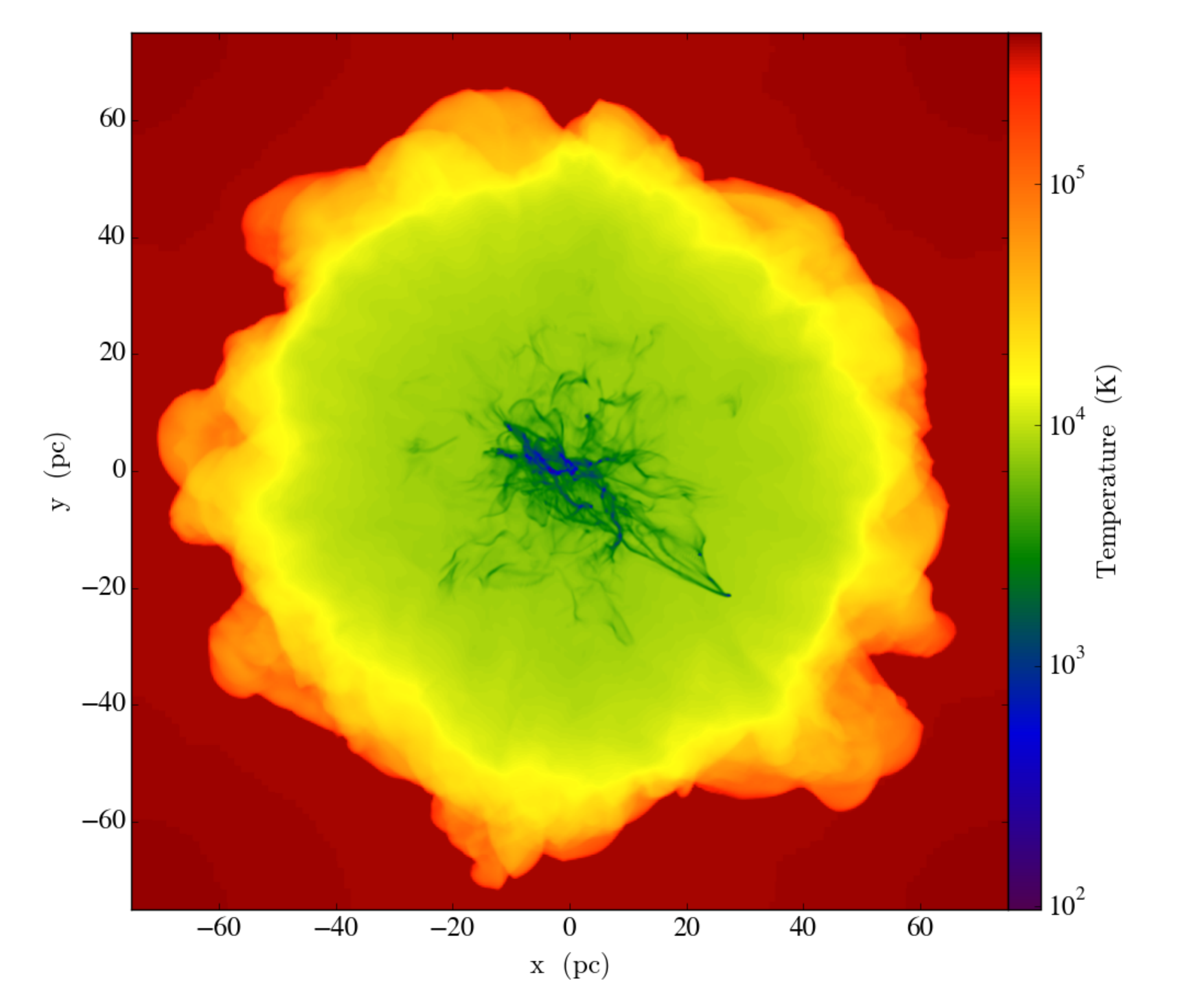}
\end{center}
\caption{\label{fig:big_projections} A projection of the surface density (left) and density-weighted temperature (right)
for the large mass and high radiative heating rate simulation at t=1.5 Myr.}
\end{figure*}

\subsubsection{Higher mass, higher radiative collapse}

Finally, to further explore the parameter space of cloud size and radiative background, we do one run with a five times larger mass,
keeping the radius constant, which results in a significant increase in the density.  Note that this means the cloud is no longer Bonner-Ebert
stable and should be gravitationally unstable to collapse even without further cooling.  Without any other change, the discussion above
makes it clear that this cloud would simply fragment before collapsing since Figure~\ref{fig:cooling_to_freefall} shows that the cooling
time to dynamical time ratio only decreases with increasing density.  Note that we keep the metallicity at $Z=10^{-2}$ $Z_{\odot}$. As we have seen, increasing the radiative heating rate would enhance the stability, forcing the evolution of the gas along the
density-temperature equilibrium curve. We test this by increasing the radiative heating rate by a factor of ten.

The result of this simulation is shown in Figure~\ref{fig:big_projections}, where, unlike the previous images, we show the density and temperature projections.
These show two things: first, as expected, the fragmentation has been delayed due to the radiative heating, giving the cloud time to globally collapse before
small-scale fragmentation occurs.  Second, these projections emphasize the rich structure visible in the full three dimensional distribution,
with clumps and filaments (these features were present in previous simulations but not as evident in the slices shown earlier). The density-temperature
profiles (not shown) are consistent with the expected density-temperature equilibrium, and the overall profiles are also consistent with the earlier simulations
which collapse globally before fragmenting.  We do note, as we will discuss in more detail below, that this experiment is, in some sense pedagogical, as we
have not attempted a self-consistent radiative transfer calculation.

%% ----------------------------------------------------------------
% 
\section{Discussion}
\label{sec:discussion}

Our numerical simulations paint a consistent picture: below a critical (low) metallicity, clouds can collapse globally before they fragment.  Radiative heating acts to boost the critical metallicity.  In the following sections, we first develop a simple analytic model which explains this result and shows how it scales with  heating rate.  Then we briefly discuss the implications before reminding readers of our approximations and how these may impact our conclusions.

\subsection{Analytic Model}

In Section~\ref{sec:basic}, we outlined our basic idea, and in the previous section, demonstrated that, in principle, the effect can be important for specific parameter choices.  Here, we try to create a {\it very} simple model of the key processes and to see how the critical cooling rate (or metallicity) scales with heating rate and cloud mass.

We begin by examining the models with radiative heating, as thermal balance results in a clear density-temperature relation.  We will then extend this to the case without (radiative) heating.  Assuming a simple power law cooling rate $\Lambda(T) = \Lambda_0 (Z/Z_0) (T/T_0)^{\alpha}$, (which is reasonably accurate below $10^4$ K, with $\alpha \approx 1$, $\Lambda_0 \approx 3 \times 10^{-26}$ erg s$^{-1}$ cm$^{3}$ at $T_0 = 10^4$ K), thermal equilibrium can be written as,
\begin{equation}
\Lambda_0 \frac{Z}{Z_0} \left( \frac{T}{T_0} \right)^{\alpha} n^2 = \Gamma n
\label{eq:thermal_eq}
\end{equation}
where $\Gamma$ is the heating rate per particle.  Conceptually fixing the temperature, this can be solved for density and used in the ideal gas equation to determine a pressure set by thermal equilibrium:
\begin{equation}
P_{\Gamma \rm eq} = \frac{\Gamma k Z_0 T_0^{\alpha}}{\Lambda_0 Z} T^{1-\alpha}
\label{eq:peq}
\end{equation}
The right panel of Figure~\ref{fig:phase_panels} (and associated discussion) shows that, before fragmentation sets in, the pressure is roughly constant, consistent with the above expression for $\alpha \approx 1$.  This fails at both low and high temperatures, but is a good approximation for a wide range of temperatures (indeed the dotted line in these figures shows a better estimate of the equilibrium pressure).

The other constraint is global hydrostatic equilibrium, which is set by the fact that the clouds generally begin with Bonner-Ebert critical initial conditions.  There are a number of ways to parameterize this central pressure, but here we choose to do so in terms of the critical Bonner-Ebert pressure:
\begin{equation}
P_{\rm cBE} = 1.4 \frac{ c_s^8}{G^3 M_{\rm BE}^2}
\label{eq:PcBE}
\end{equation}
In principle, we could rewrite $c_s$ in terms of the temperature; however, we choose not to do so in order to remind ourselves that this is a global relation (note that we could also replace $P_{\rm cBE}$ with $P_{\rm ext}$, the external pressure).

The interpretation of these two pressures can be seen conceptually with reference to the right panels of Figure~\ref{fig:phase_panels}.  For example, as the heating rate increases (or metallicity decreases), Eq.~(\ref{eq:peq}) indicates that the equilibrium pressure will increase, while the BE pressure is unchanged (this is equivalent to going from the left to right set of panels in Figure~\ref{fig:phase_panels}).  At a sufficiently high heating rate (the right column), or sufficiently low metallicity, the two curves cross only at high temperature, where the Jeans length is large and fragmentation is not allowed.  For a lower heating rate (left column) or sufficiently high metallicity, the thermal pressure allows a low-temperature solution and therefore a low Jeans length and so more rapid fragmentation.  This argument is very similar to the classic multi-phase ISM, but with the additional scale imposed by the BE sphere and the Jeans length.

The above argument indicates that there is a critical point when these two pressure coincide.  Equating our expressions for these two pressures ($P_{\Gamma \rm eq} = P_{cBE}$) gives a critical metallicity:
\begin{equation}
Z_{\rm crit} = Z_0 \frac{ k T_0^\alpha G^3}{1.4 \Lambda_0 } \frac{T^{1-\alpha}}{c_s^8} \Gamma M_{\rm BE}^2
\label{eq:crit_zl}
\end{equation}
and we can immediately read off the scaling with heating rate $\Gamma$ (linear, as expected), as well as other quantities.

In addition to the case of thermal equilibrium between radiative heating and cooling, we can explore a model in which the heating is supplied by dissipation of turbulence.  We take this heating rate as $\rho \sigma^3 / L$ , where $\sigma$ is the turbulent velocity dispersion and $L$ is the driving scale (here we assume this is given by our largest turbulent scale so that $L = 2 \pi / k_{\rm max} \approx $ 10 pc).    As before, we balance this heating with cooling, and can compute an effective equation of state.\footnote{Note that this is {\it not} the turbulent pressure, but is instead the thermal pressure that arises from turbulent heating.  A more complete description would add a turbulent pressure term, but for simplicity, here we neglect that term which we expect to be subdominant.}
\begin{equation}
P_{\rm Teq} = \frac{\sigma^3 \mu k Z_0 T_0^{\alpha}}{L \Lambda_0 Z} T^{1-\alpha}
\label{eq:Teq}
\end{equation}

Following the same logic as before, equilibrium between these pressures can be used to find a critical metallicity:
\begin{equation}
Z_{\rm crit} = Z_0 \frac{ \sigma^3 \mu k T_0^\alpha G^3}{1.4 \Lambda_0 } \frac{T^{1-\alpha}}{L c_s^8} M_{\rm BE}^2
\label{eq:crit_zt}
\end{equation}

\begin{figure}
\begin{center}
\includegraphics[width=8.5cm]{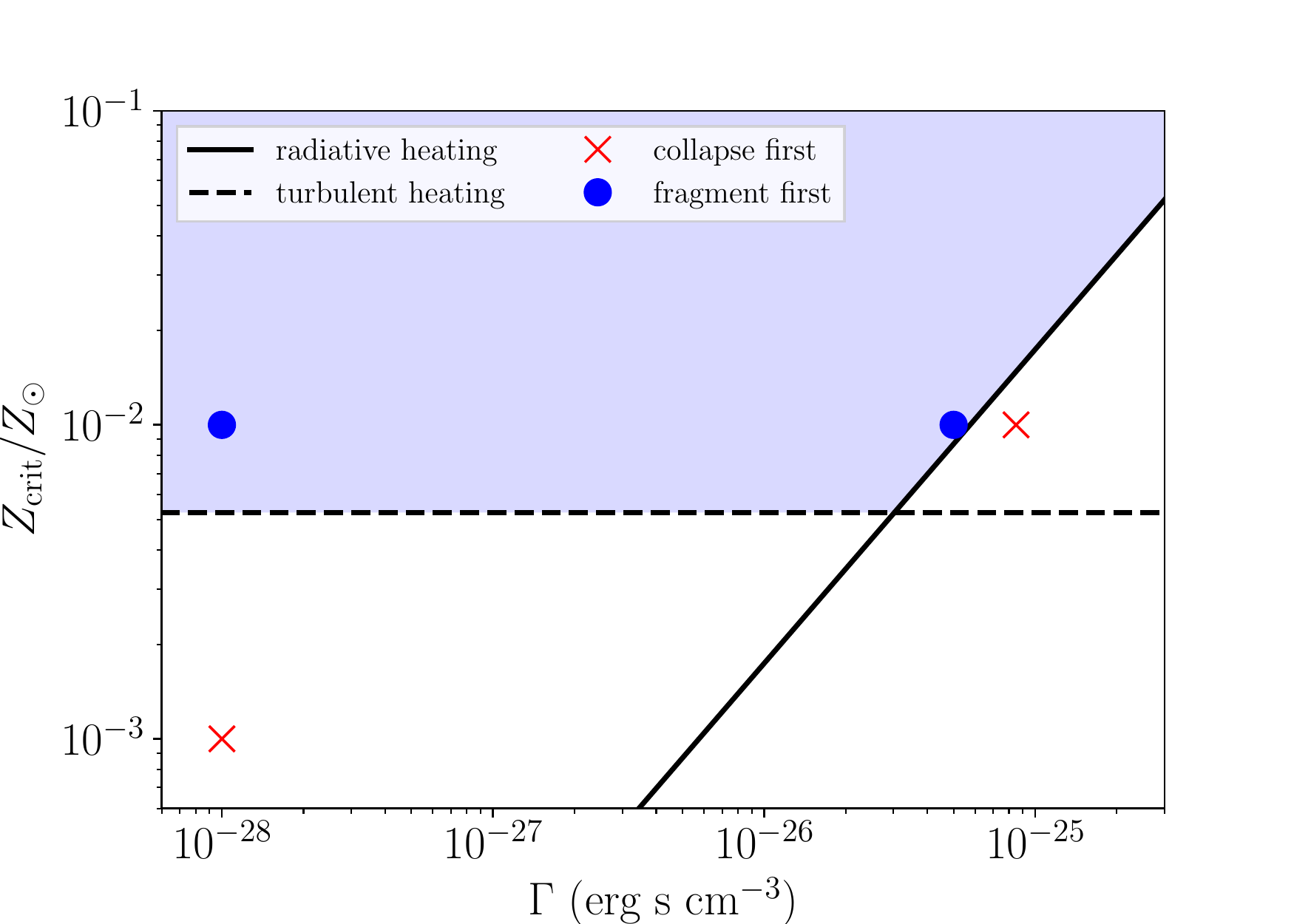}
\end{center}
\caption{\label{fig:zcrit} The critical metallicity as a function of radiative heating, indicating the dividing line between small-scale instability (above the lines, shaded blue region) and small-scale stability (below either line).  The solid line shows the relation assuming heating is radiative, while the dashed line shows the relation for turbulent heating.  Symbols correspond to numerical experiments, with crosses indicating models that collapsed before fragmenting, and circles for those that fragmented first. }
\end{figure}

In Figure~\ref{fig:zcrit}, we show these critical metallicity expressions.  The solid line shows equation~\ref{eq:crit_zl}; above this line the cooling is sufficiently rapid that the thermal pressure allows fragmentation before global collapse, while below the line radiative heating prevents fragmentation.  A similar argument holds for the turbulent-heated critical metallicity, although in this case the relation is $\sigma$ (and hence time) dependent.

Note also the dependence on the BE mass and sound speed -- in fact, this dependence is somewhat confusing, as the critical metallicity appears to increase with mass.  However, keep in mind that for a Bonner-Ebert sphere at critical stability, the external pressure actually decreases with increasing mass, and, for masses much larger than considered here, becomes unrealistically large.  As noted earlier, another way to parameterize this dependency is as $Z_{\rm crit} \propto P_{\rm cBE}^{-1}$.  This is consistent with our physical picture: a higher external pressure means that a higher equilibrium pressure can achieve balance, corresponding to a lower cooling rate and lower metallicity.  Although we have not systematically tested this scaling with simulations, our final run has a larger mass which requires a higher pressure to support, which must be offset by an increased radiative flux in order to not significantly change $Z_{\rm crit}$ (consistent with the derived scaling $Z_{\rm crit} \propto \Gamma / P_{\rm cBE}$).

\subsection{Implications}

As we have argued, the cooling efficiency may play an important role in the fragmentation of low-metallicity clouds.  Paradoxically, low metallicity (or high radiative heating) may allow the gas in near primordial clouds to cool slowly enough that global gravitational collapse precedes local gravitational collapse (at least for a time).  Our detailed simulations show that, in the absence of radiative heating, that critical metallicity is between 0.1\% and 1\% of solar metallicity.  This is well below typical ``blue" globular cluster metallicities, which are typically around a few percent \citep{West2004}.  However, we have also shown that radiative heating can boost that critical metallicity to values similar to those observed.

Of course, our calculations are merely suggestive at this point, with many approximations, as we will discuss in more detail in the next section.  However, if we take this seriously, then it does provide an interesting ``direct" explanation for the observed GC bi-modality.  On the other hand, it begs the question about what creates higher-metallicity systems.  In these cases, observations strongly suggest that rapid gas flows are required -- for example the kind of colliding flows that we expect in galaxy mergers.  In that case, the high velocities from the initial conditions (and presumably rapid cooling) permit gas to accumulate to high densities in a short period, circumventing the problem of small-scale fragmentation.  In essence, rather than allowing the $\sim 10^6$ \msun\ cloud to collapse under its own gravity, the collapse time is boosted by the velocities of the larger (galaxy-mass) halo.  

Therefore, a possible picture emerges of two modes of GC formation: one is a slow, low-metallicity mode in which the collapse timescale is the cooling time discussed above $\sim 10$ Myr; while the other is a high-metallicity, merger-mediate mode with timescale $\sim R/v_{\rm vir} \sim 1$ Myr (assuming $R \approx 50$ pc and $v_{\rm vir} \sim 200$ km/s).

\subsection{Caveats}

Here, we briefly remind readers of the many simplifications.  As we have argued throughout, these calculations are intended to more suggestive explorations than realistic predictions.  

First, the initial conditions are quite simplistic, with a Bonner-Ebert sphere and some imposed turbulence.  We do not expect the results to strongly depend on the details of the turbulence; however we have not really discussed how such initial conditions might arise.  One possibility is that colliding flows in low $v_{\rm vir}$ halos shock heat and cool down to $10^4$ K (but not below, due to the break in the cooling rate), settling into the equilibrium configuration envisioned.  More realistic (future) work would explore colliding flows or colliding clouds to better understand the transition between the fast and slow modes discussed above.  Ideally, the initial conditions would be drawn from large-scale cosmological simulations which self-consistently model the gas motions within realistic dark-matter halos and the build-up of metals and external radiation fields (but cannot follow the small-scale fragmentation modeled here).

In addition, our radiative and chemical model is somewhat idealized.  Although the \textsc{grackle} cooling we adopt is based on Cloudy \citep{Cloudy2017} and so, radiative processes aside, quite accurate, our treatment of a uniformly constant radiative heating rate fails to take into account the full complexity of more realistic treatments.  Our uniform rate is best explained by a far-UV photo-electric heating source; however, the low metal content we assume implies a similarly low (or lower) dust content, reducing the efficiency of photoelectric heating.  For example, a heating rate for run 4 of $8.5 \times 10^{-25}$ erg s$^{-1}$ cm$^{-3}$ corresponds approximately to the emission from a nearby star-forming region 300 pc away producing stars at the rate of about 1 M$_{\odot}$/yr.    Finally, we do not include H$_2$ cooling which may be important \citep{Glover2014}, although a strong Lyman-Werner background will help to photo-dissociate the molecules.

There are also a range of other physical processes that we do not include which may play a role in such systems, of which the absence of magnetic fields may be the most glaring.  Finally, we explore the fragmentation of the clouds, but do not attempt to model the formation of individual stars, nor their feedback effects.  Previous work that has examined the impact of feedback finds that in order to form bound clusters, the efficiency of star formation must be high \citep{Bastion2006}, a result which helps to motivate our requirement that fragmentation must not occur until the gas density is very high (so that fragmentation completes on a short timescale, before feedback can impact the cluster evolution).

%% ----------------------------------------------------------------

\section{Summary}

We have carried out simulations of low metallicity (but not primordial) gas clouds in order to better understand  how such collapse proceeds.  This is motivated in part by a suggestion that the low-metallicity (blue) population of globular clusters form preferentially from such low metallicity gas due to the possibility that the dynamical time is shorter than the cooling time in these clouds.  In particular, we carried out a set of simulations of Bonner-Ebert stable clouds with densities of 10-100 cm$^{-3}$, radii of about 50 pc, and temperatures just below $10^4$ K, with the following results.

\begin{enumerate}

\item In the absence of radiative heating, there is a critical metallicity between $Z=0.001$ Z$_\odot$ and $Z=0.01$ Z$_\odot$, below which the cloud first collapses globally before fragmenting.  For metallicities larger than this critical value, the opposite occurs, and the cloud first fragments.  This is shown most clearly in Figure~\ref{fig:number_density_panel}.  The low metallicity runs result is a fragmenting cloud with properties (size and mass) reminiscent of present-day globular clusters; however, we note that this critical metallicity is lower than the typical values seen even in low-metallicity globular clusters.

\item Adding a radiative heating source (perhaps due to photo-electric heating although we have not included radiative transfer) changes the evolution.  It establishes an equilibrium density-temperature curve and the gas ends up close to this curve in the simulations.  Because of the temperature dependence of the radiative cooling at low temperatures, this gas is nearly at constant pressure.  We find that, if this thermal-equilibrium set pressure is lower than the hydrostatic pressure of the cloud, rapid fragmentation can occur; while if it is higher, then the cloud is largely stable.  If the stable branch can then gradually collapse (either because the radiative heating rate declines or the cloud is driven globally gravitationally unstable due to cloud-cloud collisions or ongoing accretion), then high-density GC-like formation can occur.

\item For the radiative-heating case, we developed a simple analytical model based on the relative magnitude of the two pressures described above (the thermal-equilibrium pressure and the hydrostatic pressure), with the outcome either fragmentation or (local) stability.  This led again to the identification of a critical metallicity, but now one that depends on the radiative heating rate.  This is shown in Figure~\ref{fig:zcrit}.

\end{enumerate}

Future steps would be to (i) improve the microphysical model with a more realistic heating and cooling model; (ii) better link the initial conditions to cosmological conditions in the high-redshift forming halos that host these gas clouds; and (iii) model the fragmentation to predict the initial mass function of such clouds, rather than a simple fragmentation or no-fragmentation criterion.

%% ----------------------------------------------------------------
% 
\section*{Acknowledgments}

We thank the referee, Paul Clark, for comments which helped improve the presentation of this paper.
We acknowledge support from NSF grants AST-1312888 and AST-1615955, and NASA grant NNX15AB20G, as well as computational resources from NSF XSEDE,
and Columbia University's Yeti cluster.  We thank Simon Glover for useful conversations.  Simulations were run on the Texas Advanced Computing Center Stampede supercomputer and were performed
using the publicly-available Enzo code (http://enzo-project.org) and analyzed with the yt package \citep{Turk2011}. Enzo is the product of a
collaborative effort of many independent scientists from numerous institutions around the world. Their commitment to open science has helped make
this work possible. The Flatiron Institute is supported by the Simons Foundation. 

\bibliography{mn-jour,gc_paper}
%
%\label{lastpage}
\end{document}